\documentclass[12pt]{iopart}
\usepackage[english]{babel}
\usepackage{epsfig}
\usepackage{changebar}
\usepackage{graphicx}
\begin{document}
\renewcommand{\figurename}{{\bf Fig.}}
\renewcommand{\tablename}{{\bf Tab.}}
\title{Conditions driving chemical freeze-out}

\author{A.~Tawfik\footnote[5]{tawfik@physik.uni-bielefeld.de}}

\address{University of Bielefeld, P.O.~Box 100131, D-33501~Bielefeld, 
              Germany}

\begin{abstract}
We propose the entropy density as the thermodynamic condition driving best
the chemical freeze-out in heavy-ion collisions. Taking its value from
lattice calculations, we find that it is excellent in reproducing the
experimentally estimated freeze-out parameters. The two characteristic
endpoints in the freeze-out diagram are reproduced as well. 
\end{abstract}

\pacs{12.40.Ee,12.40.Yx,05.70.Ce}

\maketitle

\section{\label{sec:1}Introduction}

At critical temperature $T_c$, the hadronic matter is conjectured to
dissolve into quark-gluon plasma (QGP). Reducing the QGP temperature
disposes hadronization and the interacting system goes into chemical
equilibrium and finally the produced particles freeze out. Below $T_{ch}$,
thermal equilibrium takes place and the compositions of matter turn to
confine into 
hadrons. The best way to determine the freeze-out parameters is to combine
various particle ratios in order to obtain a window in the $T_{ch}-\mu_B$
diagram compatible with the experimental results. $\mu_B$ is the
baryo-chemical potential. The question we intent to answer is: what is the
universal condition describing the freeze-out
parameters?~\cite{Braun-Munzinger:1996mq,Cleymans:1999st}

{\it Without energy input the chemical reactions always proceed toward
chemical equilibrium, i.e. balancing particle-absorption and
-production}. To study the consequences of this equilibrium, we recall the
equilibrium constant ${\cal C}_{ch}$. In a chemical reaction like $aA + bB
\rightarrow cC + dD$, ${\cal C}_{ch}$  can be calculated according to
the ''law of mass action''~\cite{Greiner:2004vm,Tawfik:2004vv}   
\begin{eqnarray}
{\cal C}_{ch} &=& \left. \left([C]^c_{ch} \;
  [D]^d_{ch}\right)\right/\left([A]^a_{ch}\;  [B]^b_{ch}\right),
  \label{eq:chconst} 
\end{eqnarray}
where $(A,B)$ and $(C,D)$ refer to the reactants and products,
respectively. $(a,b)$ and $(c,d)$  are the corresponding concentrations. 
In heavy-ion collisions, the decay channels  are - in structure -  similar
to the above chemical reaction. The backward direction can be viewed as
annihilation/absorption processes. Analogy to Eq.~(\ref{eq:chconst}) we
deal so far with {\tt one specific type of hadron interactions}. The hadron
system and its approach towards the chemical equilibrium are allowed to
have many dynamical processes. The most important ones are 
particle-absorption and -production. The different mechanisms that drive
the system towards the chemical equilibrium are to be taken into
consideration by summing over all hadron resonances. 

The total free energy of the system $\delta  G$ can be used to determine
the likely direction. When the reactant and product concentrations are 
given, then for an ideal gas (no enthalpy change), $\delta  G = \delta
G^0 + T \ln ({\cal C}_{ch})$. At equilibrium free energy gets minimum
($\delta G$ vanishes) and pressure is constant. Then ${\cal C}_{ch} =
\exp(-\delta G^0/T)$. $\delta G^0=\delta E -T\delta S$ is the difference of
free energies of products and reactants at standard state, i.e. at  $T\neq
T_{ch}$.    
\begin{eqnarray}
{\cal C}_{ch} &\approx&\exp(-\delta s - \mu\,\delta n/T), \label{eq:chconst3}
\end{eqnarray}
At equilibrium, the entropy gets maximum. The change in particle number is
minimum. Therefore, the equilibrium constant is mainly influenced by
entropy change.

\section{\label{sec:2}Thermodynamic conditions}

In the following, we list some thermodynamic
  expressions for one particle and its anti-particle in Boltzmann limit.   
\begin{eqnarray}
n(T,\mu_B) &=& \frac{g}{\pi^2} T
             m^2 K_2\left(\frac{m}{T}\right) \;
             {\mathbf \sinh\left(\frac{\mu_B}{T}\right)}, \label{eq:n1} \\
s(T,\mu_B) &=& \frac{g}{\pi^2}
             m^2 \left[m K_3\left(\frac{m}{T}\right)
             \; {\mathbf \cosh\left(\frac{\mu_B}{T}\right)}
             \right. \nonumber \\
            & &\left.\hspace*{8mm} - \mu_B K_2\left(\frac{m}{T}\right) 
             \; {\mathbf \sinh\left(\frac{\mu_B}{T}\right)}\right],
             \label{eq:entr}  \\
\frac{\epsilon(T,\mu_B)}{n(T,\mu_B)} &=& \left[3T + m
             \frac{K_1\left(\frac{m}{T}\right)}{K_2
             \left(\frac{m}{T}\right)}\right] \; {\mathbf
             \coth\left(\frac{\mu_B}{T}\right)},    \label{eq:eOvern1} 
\end{eqnarray}
where $g$ is the spin-isospin degeneracy factor and $K_i$ are the modified
Bessel functions. They will be summed over all resonances taken into
account. These quantities are related to the consequences of chemical
equilibrium via Eq.~(\ref{eq:chconst3}). In this letter, full quantum
statistics has been properly taken into account.  
\begin{eqnarray}
\ln {\cal Z}(T,\mu_B) &=& \frac{g\,V}{2\pi^2} \int_{0}^{\infty}
           k^2 dk  \ln\left[1 \pm\,
           e^{\frac{\mu_B-\varepsilon}{T}}\right],
           \label{eq:lnz} \hspace*{5mm} 
\end{eqnarray}
where $\varepsilon=(k^2+m^2)^{1/2}$ is the single-particle energy and
$\pm$ stand for bosons and fermions, respectively. We apply the hadron
resonance gas model (HRGM)~\cite{Karsch:2003vd, Karsch:2003zq,
  Redlich:2004gp, Tawfik:2004sw} in order to study the conditions driving
the freeze-out. All observed resonances up to mass 
$2\;$GeV are included. The particle decays are entirely left away. We use
grand canonical ensemble and full quantum statistics. Corrections due to
van~der~Waals repulsive interactions and the excluded volume 
have not been taken into account.

\begin{figure}[thb]  
\centerline{\includegraphics[width=8.cm]{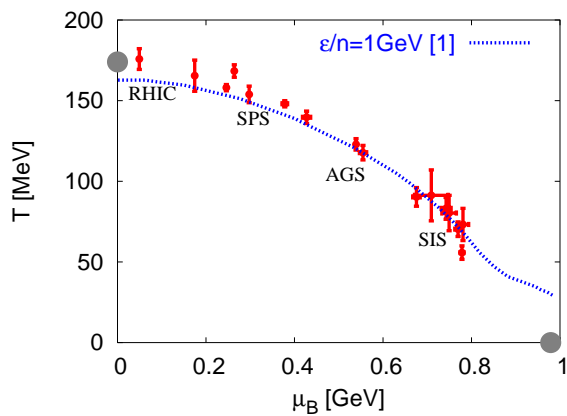}}  
\caption{\label{Fig:1}\footnotesize
  Freeze-out curve under the condition of constant energy per
  particle~\cite{Cleymans:1999st}. The points are the parameters
  taken from indicated accelerators.} 
\end{figure}

\begin{figure}[thb] 
\centerline{\includegraphics[width=8.cm]{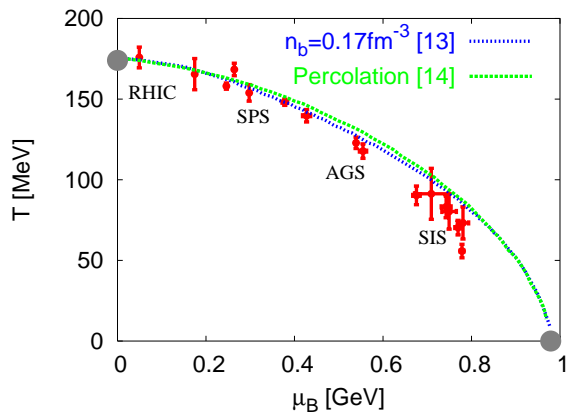}} 
\caption{\label{Fig:2}\footnotesize
  Freeze-out curve under the
  condition~\hbox{$n_B=0.17\;$fm$^{-3}$}~\cite{Braun-Munzinger:2001mh}. In 
  calculating the short-dashed curve~\cite{Magas:2003wi}, the percolation
  theory has been taken into account. }     
\end{figure}

Before we report the results, we remind of the two characteristic points of
the freeze-out diagram; One at $T_{ch}\neq0$ and
\hbox{$\mu_B=0$} and the other at $T_{ch}=0$ and very large
$\mu_B$. Localizing the first point has been the subject of different
experimental studies~\cite{Becattini:2002nv}. It has been found that
\hbox{$T_{ch}(\mu_B=0)\approx 174\;$MeV}~\cite{Braun-Munzinger:2003zz}. The
lattice estimation for the deconfinement temperature  is
\hbox{$T_{c}(\mu_B=0)=173\pm8\;$MeV}~\cite{Karsch:2001cy}. This  
implies that the deconfinement and freeze-out lines seem to be coincident at
low $\mu_B$. For the second point, we are left with applying effective
models. As $T\rightarrow0$, the nucleons in the hadron gas get
dominant. Applying Fermi statistics, Eq.~\ref{eq:raioT0}, we find that 
the chemical potential corresponding to the normal nuclear density,
\hbox{$n_0\approx0.17\;$fm$^{-3}$} is \hbox{$\mu_{ch}\approx0.979\;$GeV}.   
This value can slightly be different according to the initial
conditions~\cite{Fraga:2003uh}. 

A proposal to describe the freeze-out parameters at different incident 
energies has been reported in Ref.~\cite{Cleymans:1999st}. The authors
started from phenomenological observations at GSI/SIS energy that
$\epsilon/n\approx1\,$GeV. They have applied Boltzmann approximations in
calculating $\epsilon/n$, Eq.~(\ref{eq:eOvern1}), and analytically handled
the hadron resonance gas at $T\approx50\,$MeV ($\mu_B\approx0.8\,$GeV),
i.e. the freeze-out parameters at GSI/SIS energy, as a Fermi gas of degenerate
nucleons, Eq.~\ref{eq:raioT0}. At high $T$ and small $\mu_B$, the authors
assumed that the pions and rho-mesons get dominant. The baryons are entirely
neglected. Applying Eq.~(\ref{eq:eOvern1}) with the 
effective mass $m=m_{\pi}+m_{\rho}$ they got almost the same value for the
ratio $\epsilon/n$~\cite{Cleymans:1999st}. In Fig.~\ref{Fig:1}, we use HRGM
in order to map out $T-\mu_{ch}$ freeze-out diagram according to this
condition. In calculating $n$ and $\epsilon$, we take into account full
quantum statistics and all observed resonances up to mass $2\,$GeV.

In Ref.~\cite{Braun-Munzinger:2001mh}, {\it total} baryon
number density $n_B$ was imposed to interpret the chemical freeze-out
curve. 
For this scenario, the baryon-baryon and baryon-meson interactions were
assumed to drive the chemical equilibrium. For the value 
$n_B=0.12\;$fm$^{-3}$, two-third the normal nuclear density, it has been
argued that it depends on the correction due to van~der~Waals repulsive
interactions. In Fig.~\ref{Fig:2}, we plot HRGM results under the condition
\hbox{$n_B=0.17\;$fm$^{-3}$}, the normal nuclear density. We use this
value, since we entirely leave away all corrections. We find that our
results suggest that the assumption of baryon-baryon and baryon-meson and
the repulsive interactions are not well-founded. Nevertheless, we find that
the value \hbox{$n_B=0.17\;$fm$^{-3}$} is satisfactorily able to 
reproduce the two endpoints of freeze-out diagram.

Another model we consider is in the framework of percolation
theory~\cite{Magas:2003wi}. At \hbox{$T=0$}, the freeze-out occurs when 
the nucleons no longer form interconnected matter. The corresponding density 
is found to be \hbox{$\approx0.17\;$fm$^{-3}$} and consequently,
$\mu_B=0.979\,$GeV. At $\mu_B=0$, it has been found that 
\hbox{$T_{ch}\approx175\;$MeV} for
$\gamma_s=0.5$~\cite{Magas:2003wi}. $\gamma_s$ gives the strangeness
saturation. 
The results are shown in Fig.~\ref{Fig:2}.  

We can so far conclude that the last two
models~\cite{Braun-Munzinger:2001mh, Magas:2003wi} are able to reproduce
the two endpoints. Both apparently overestimate the
freeze-out parameters at BNL/AGS and GSI/SIS
energies. Model~\cite{Cleymans:1999st} describes well the 
freeze-out parameters at low energy. It slightly underestimates the
parameters at BNL/RHIC and CERN/SPS energies. Its largest discrepancy is at
energies lower than GSI/SIS. At $\mu_B=0.979\;$GeV, which as given
above corresponds to $n_0$, we find that the freeze-out 
under the condition $\epsilon/n=1\,$GeV occurs at $T_{ch}\sim35\;$MeV!
Furthermore, extrapolating the resulting curve to abscissa results in
particle number density $25-30\, n_0$. According to Eq.~\ref{eq:raioT0},
the ratio $\epsilon/n$ at $T=0$ and $\mu_B=0.979\;$GeV equals $2.89\,$GeV. 
\begin{eqnarray}
\frac{\epsilon(\mu_B)}{n(\mu_B)} &=& 9\, g\, m^4\;\frac{
       \frac{\mu_B}{m}  
       \sqrt{\frac{\mu_B^2}{m^2}-1}
       \left(\frac{\mu_B^2}{m^2}-\frac{1}{2}\right) - \frac{1}{2}
       \ln\left(\frac{\mu_B}{m} +
       \sqrt{\frac{\mu_B^2}{m^2}-1}\right)}{ 
       8\left(\mu_B^2 -m^2\right)^{3/2}}. \hspace*{6mm}
\label{eq:raioT0}
\end{eqnarray}
This relation is valid for fermions and therefore, we drop
out the exponent $\pm1$~\cite{Cleymans:1985wb}.

\begin{figure}[thb] 
\centerline{\includegraphics[width=8.cm]{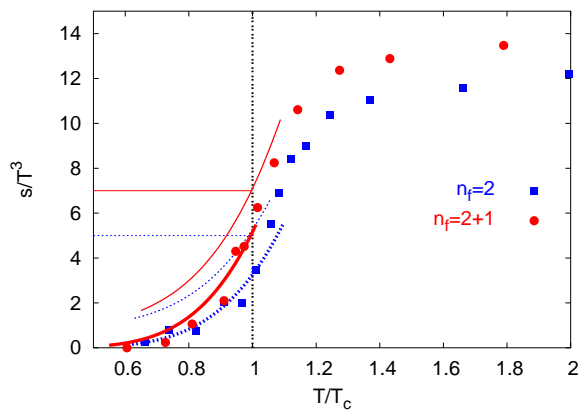}} 
\caption{\label{Fig:3}\footnotesize
Lattice QCD results on entropy density normalized to
$T^3$ for $2$ and $2+1$ quark flavors at $\mu_B=0$. The thin curves give 
HRGM results for physical masses. The results for re-scaled resonance
masses are given by the other two curves. 
}    
\end{figure}
\vspace*{-3mm}

\begin{figure}[thb] 
\centerline{\includegraphics[width=8.cm]{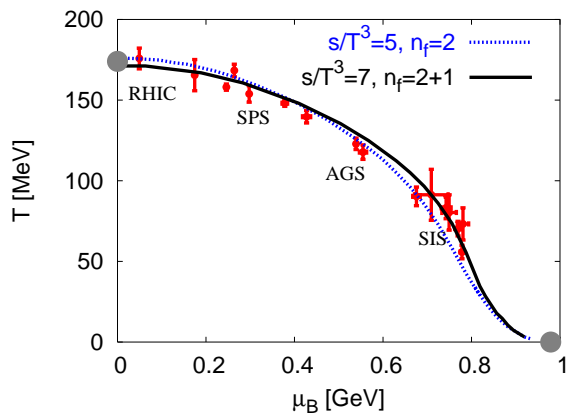}} 
\caption{\label{Fig:4}\footnotesize
  The freeze-out curve under the condition of constant
  $s/T^3$. For non-strange resonances, $s/T^3=5$ and for all resonances,  
  $s/T^3=7$ (Fig.~\ref{Fig:3}). The two solid circles as well as the
  experimentally estimated points are very well reproduced.}  
\end{figure}

\begin{figure}[thb] 
\centerline{\includegraphics[width=8.cm]{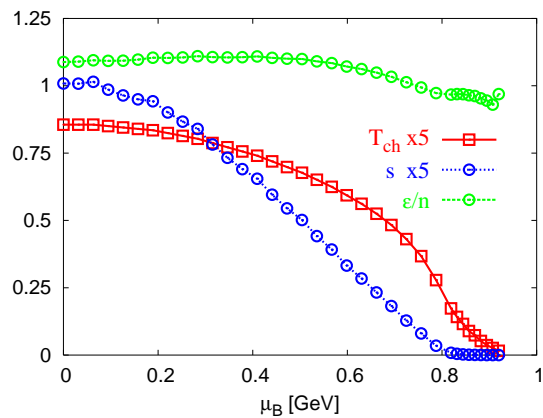}}
\caption{\label{Fig:5}\footnotesize
  $T_{ch}$, $s$ and $\epsilon/n$ are calculated along the freeze-out
  curve. We 
  notice that $s$ decreases much faster than $T_{ch}$, so that at very large
  $\mu_B$ the condition $s/T^3=7$ is no longer valid. The ratio
  $\epsilon/n=1\,$GeV does not remain constant along $\mu_B$-axis.}  
\end{figure}

\section{\label{sec:3}Entropy density at chemical freeze-out}

As mentioned above, our condition for chemical equilibrium in heavy-ion
collisions is the entropy density, Eq.~\ref{eq:chconst3}~
\cite{Tawfik:2004vv}. For vanishing free energy, the equilibrium entropy
gives the amount of energy which can't be used to produce additional
work. In this context, the entropy can be defined as the degree of sharing and
spreading the energy inside the system. The way of distributing the energy
is not just an average value, but rather the way that controls the chemical
equilibrium.  

As $T\rightarrow0$ and $\mu_B\neq0$, the thermodynamic quantities are given
in Eq.~(\ref{eq:raioT0}). In this limit, the entropy density vanishes and
HRGM is no longer applicable. At $\mu_B=0$, the system becomes
meson-dominant. In this case, the entropy is
finite~\cite{Becattini:2002nv,Karsch:2001cy}. What is the 
entropy that characterizes the freeze-out at 
finite $T$ and $\mu_B$? As mentioned above, the deconfinement transition is
coincident with the freeze-out at small $\mu_B$. We therefore
rely on the lattice calculations~\cite{Karsch:2003vd, Karsch:2003zq,
  Redlich:2004gp,  Karsch:2000kv}.  
In Fig.~\ref{Fig:3}, we plot the lattice results on $s/T^3$ vs. $T/T_c$ at
$\mu_B=0$~\cite{Karsch:2000kv,Karsch:2003zq} for $n_f=2$ and $n_f=2+1$
quark flavors. To compare with the
lattice calculations in which very heavy quark masses are used, the resonance
masses have to be re-scaled to values heavier than the physical
ones~\cite{Karsch:2003vd,Karsch:2003zq}. The results with the physical 
masses are given by the thin curves. The two horizontal lines point at the
value of $s/T^3$ at corresponding $T_c$. For physical masses,
we find that $s/T^3=5$ for $n_f=2$ and $s/T^3=7$ for $n_f=2+1$. The  
normalization with respect to $T^3$ should not be connected with massless
ideal gas. Either the resonances in HRGM or the quarks on lattice are
massive.

At given $\mu_B$, we calculate $T_{ch}$ according to constant
$s/T^3$. The results are plotted in Fig.~\ref{Fig:4}. The dotted curve
represents $n_f=2$ results, ($s/T^3=5$). The solid curve gives $n_f=2+1$
results, ($s/T^3=7$). We 
find that the characteristic endpoints as well as all experimentally
estimated freeze-out parameters are very well reproduced. Comparing with
Fig.~\ref{Fig:1} and Fig.~\ref{Fig:2}, one finds 
that our results fit best freeze-out parameters.
In  Fig.~\ref{Fig:5}, we plot $\epsilon/n$, $T_{ch}$ and $s$ in dependence
on $\mu_B$ along the freeze-out curve. We notice that as $\mu_B$ increases,
both $s$ and $T_{ch}$ decrease, too. We also find that $s$ decreases much
faster than $T$. The ratio $s/T^3$ becomes greater than $7$ at very
large $\mu_B$. In this limit, HRGM is no longer applicable and the numerics
terminates. At very large $\mu_B$, {\it thermal} entropy $s$ is expected to
vanish, since it becomes proportional to $T$ (third law of
thermodynamics). The {\it quantum} entropy~\cite{Miller:2003ha,
  Miller:2003hh, Miller:2003ch, Miller:2004uc, Hamieh:2004ni,
  Miller:2004em} is entirely disregarded here.  We also find that  
$\epsilon/n$ does not seem to remain constant along the freeze-out curve as
assumed in~\cite{Cleymans:1999st}. Since HRGM is not applicable in the limit
$T=0$, we apply Fermi statistics, Eq.~\ref{eq:raioT0}, as given above.

\section{\label{sec:4}Conclusion}

We reviewed the conditions suggested to describe the freeze-out
parameters. We used HRGM in order to map out the freeze-out curve according
to these conditions. Full quantum statistics has been properly taken into
account. We compared the results of the models proposed and check their
abilities in reproducing the experimentally estimated freeze-out parameters
and the characteristic endpoints (Fig.~\ref{Fig:1} and
Fig.~\ref{Fig:2}). We found that there are different constraints in
reproducing the endpoints and fitting the parameters.   

We proposed the entropy density $s$ as the thermodynamic condition driving
the chemical freeze-out, Eq.~\ref{eq:chconst3}. Taking its value from
lattice QCD simulations at $\mu_B=0$ and assuming that $s$ normalized to
$T^3$ remains constant in the entire $\mu_B$-axis, we obtained the results
shown in Fig.~\ref{Fig:4}. The freeze-out parameters
$T_{ch}$ and $\mu_B$ are very well described under this condition. The two
characteristic endpoints are also reproduced.    
We conclude that {\sf the given ratio $s/T^3$ characterizes very
well the final states observed in all heavy-ion experiments}. Increasing
the incident energy leads to increasing particle yields. The
production rates of particles decrease exponentially with their
masses. This phenomenological observations are regarded in the way, that
increasing energy/temperature is considered by including heavy
resonances. Changing energy with changing particle number is given by
chemical potential $\mu_B$. The amount of energy which produces no
additional work, i.e. vanishing free energy, is the entropy at chemical
equilibrium.  \\  

I thank Rudolf~Baier, Tamas~Biro, Rajiv~Gavai, Krzysztof~Redlich and
Helmut~Satz for useful discussions.  

\hspace*{1cm}


\end{document}